\documentclass[fleqn,12pt,twoside]{article}
\usepackage{espcrc1}

\usepackage{graphicx}
\usepackage[figuresright]{rotating}

\newcommand{\AmS}{{\protect\the\textfont2
  A\kern-.1667em\lower.5ex\hbox{M}\kern-.125emS}}

\title{The Neutrino Signal in Stellar Core Collapse and Postbounce Evolution}

\author{
M. Liebend\"orfer\address[UTK]{Department of Physics \& Astronomy, University of Tennessee, Knoxville TN 37996}\address[ORNL]{Physics Division, Oak Ridge National Laboratory, Oak Ridge TN 37831}\address[CITA]{Canadian Institute for Theoretical Astrophysics, Toronto ON M5S 3H8},
A. Mezzacappa\addressmark[ORNL],
O.~E.~B. Messer\addressmark[UTK]\addressmark[ORNL],
G. Martinez-Pinedo\address[UBA]{Department of Physics \& Astronomy, University of Basel, Basel CH 4056}\address[IEEC]{Institut d'Estudis Espacials de Catalunya, Barcelona E 08034},
W.~R. Hix\addressmark[UTK]\addressmark[ORNL],
and
F.-K. Thielemann\addressmark[UBA]
}
       
\begin{document}

% typeset front matter
\maketitle

\begin{abstract}
General relativistic multi-group and multi-flavor Boltzmann neutrino
transport in spherical symmetry adds a new level of
detail to the numerical bridge between microscopic nuclear and weak
interaction physics and the macroscopic evolution of the astrophysical
object. Although no supernova explosions are obtained, we investigate
the neutrino luminosities in various phases of the postbounce evolution
for a wide range of progenitor stars between 13 and 40 solar masses.
The signal probes the dynamics of material layered in and around the
protoneutron star and is, within narrow limits, sensitive to improvements
in the weak interaction physics. Only changes that dramatically exceed
physical limitations allow experiments with exploding models. We discuss
the differences in the neutrino signal and find the electron fraction
in the innermost ejecta to exceed \( 0.5 \) as a consequence of thermal
balance and weak equilibrium at the masscut.
\end{abstract}

\section{Modeling core collapse supernovae on the computer}

The traditional scientific experiment may consist of four
parts: the preparation of the ingredients, the setup and conduct
of a measurement, and a theory to interpret the results.
The measurement and equipment present the well defined
link between reality and theory. Computer simulations in astrophysics
can take a similar r\^ole. In this case, ingredients are based on
theory and the outcome is linked to reality in form of astrophysical
observations. For example, supernovae will be reproduced
in full detail when all important ingredients are in place. There
are, however, lively discussions over which these ``important ingredients''
are. Self-consistent spherically symmetric simulations try to address
a neutrino-driven supernova mechanism \cite{Wilson_85}.
Recently, neutrino transport has been brought close to perfection in the
solution of the dynamic Boltzmann transport equation \cite{Rampp_Janka_00}
and general relativity has been included with ``standard'' nuclear
and weak interaction physics \cite{Bruenn_DeNisco_Mezzacappa_01}
to form the current state-of-the-art in spherically symmetric simulations,
where tightly coupled radiation hydrodynamics is followed in three
dimensions: space, neutrino propagation direction, and neutrino energy.
All recent simulations have corroborated the qualitative result that
neutrinos don't drive a supernova in spherical symmetry. There is
broad agreement that convectional instabilities behind the outwards
propagating shock increase the heating efficiency. While multi-dimensional
simulations with neutrino transport approximations have delivered encouraging
results \cite{Herant_et_al_94},
more recent calculations, combining one-dimensional
energy dependent transport with two-dimensional hydrodynamics, again
failed to explode the stellar envelope \cite{Mezzacappa_et_al_98}.
No doubt that convection is an important ingredient, but there might
be more. Convection in the protoneutron star \cite{Wilson_Mayle_93},
significant rotation, differences between 2D
and 3D simulations \cite{Fryer_Heger_00}, and magnetic fields \cite{MacFadyen_Woosley_99}
are permanent entries on the to do list of future investigations with
accurate neutrino transport.

\section{Sources and impact of neutrino luminosities\label{section_neutrino_sources}}

In Fig. (1),
\begin{figure}[p]
\includegraphics[width=0.95\textwidth]{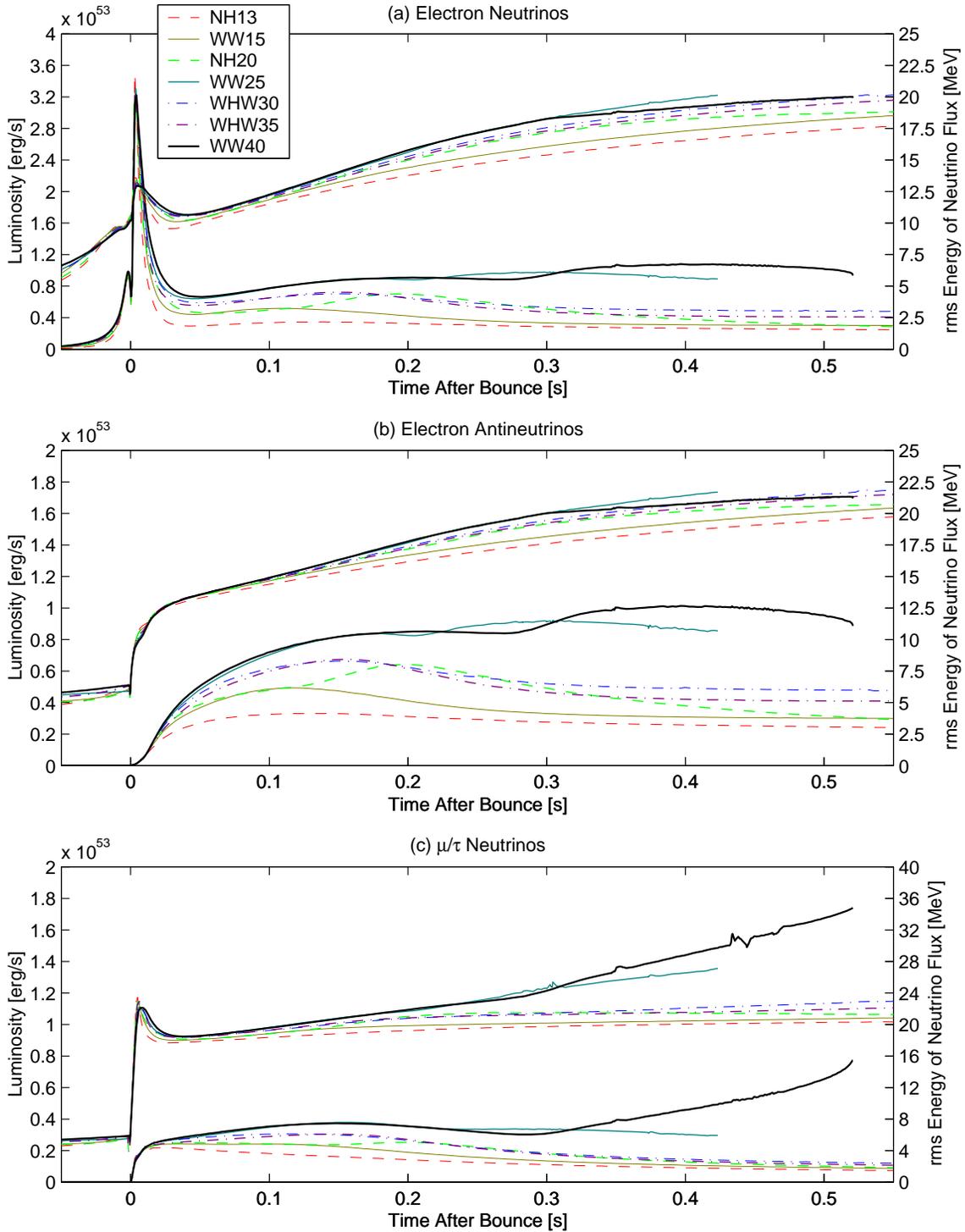}
\caption{Neutrino luminosities and rms energies of the neutrino flux are shown
as a function of time in the evolution of various progenitor models. The upper
bundle of curves belongs to the right axis, the lower bundle to the left axis.
The quantities are sampled at \protect\( 500 \protect\) km in the comoving
frame.}
\end{figure}
we present the luminosities and rms energies in simulations
issued from different progenitor stars. From \cite{Nomoto_Hashimoto_88}
we evolved a \( 13 \) M\( _{\odot } \) and a \( 20 \) M\( _{\odot } \)
star; from \cite{Woosley_Weaver_95} we evolved a \( 15 \) M\( _{\odot } \),
a \( 25 \) M\( _{\odot } \), and a \( 40 \) M\( _{\odot } \) star;
and from \cite{Woosley_Heger_Weaver_02} a \( 30 \) M\( _{\odot } \)
star, and a \( 35 \) M\( _{\odot } \) star with zero initial metallicity.
The simulations with our code {\sc agile-boltztran} \cite{Mezzacappa_Messer_99}
are based on the Lattimer-Swesty equation of state
and ``standard'' weak interactions \cite{Lattimer_Swesty_91}.
The electron neutrino luminosities are shown in Fig. (1a).
The luminosities rise during collapse. Immediately
after bounce, a shock is formed at an enclosed mass of \( 0.53 \)
M\( _{\odot } \). It propagates a density jump outwards that compresses
neutrino diffusive material to neutrino opaque matter. This
leads to the \( 4 \) ms short dip visible in the luminosities right
after bounce. As soon as the postshock densities become neutrino
transparent, a neutrino burst is released by immediate electron captures
on the shock-dissociated nucleons. The evolution up to the neutrino
burst is similar in all models because of a self-regulation mechanism
during collapse \cite{Messer_et_al_02}: For larger electron fractions
many more protons are available for electron capture and the
electron fraction decays more quickly. If electron capture on
nuclei were to dominate, this self similarity might not occur.
After this initial phase,
the neutrino luminosities of the different models open up to individual
signatures. The luminosities, as a function of time, reflect
the accretion rates (i.e. the density profiles in the outer layers
of the progenitor stars) and the gravitational potential at the surface
of the protoneutron star \cite{Liebendoerfer_et_al_02}.
In order to exploit the details of our calculations, we append Fig.
(2) for the self-guided contemplation of the interested reader.
\begin{figure}[t]
\includegraphics[width=0.9\textwidth]{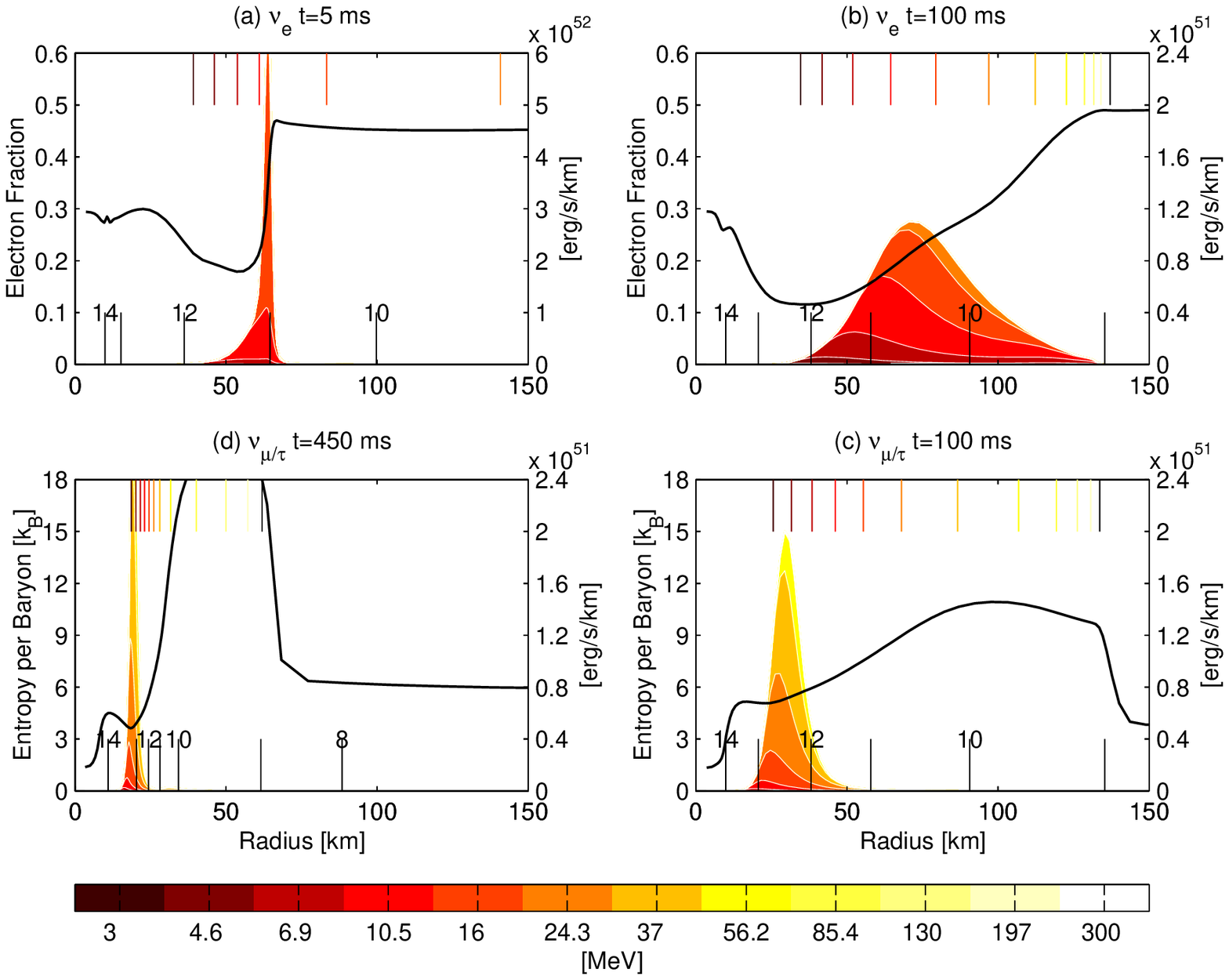}
\caption{The production site of escaping neutrinos in WW15 is shown
as a function of radius in units of {[}erg/s/km{]}.
The graphs also show profiles of the electron fraction and matter entropy.
Markers in the lower part of the graphs indicate the logarithmic density
in g/cm\protect\( ^3 \protect\). In the upper part, they indicate the
scattering-dominated neutrinospheres for rising energy groups from left to right
according to the legend.
Note the sharp profile of the neutrino burst and the broad emission
region at \protect\( 100 \protect\) ms after bounce.
Cooling by \protect\( \mu/\tau \protect\) neutrino emission shrinks
at late times to a very narrow radius interval.}
\end{figure}

The traditional analysis of the neutrino-driven supernova mechanism
refers to neutrinospheres, the location where neutrinos have an
optical depth of \( 2/3 \). Between the neutrino
sphere and the shock, neutrino emission and neutrino absorption takes
place in the fully dissociated, shock-heated, inwards drifting material.
A ``gain radius'' separates an inner cooling region, where neutrino
emission dominates, from an exterior heating region, where neutrino
absorption dominates. However,
the concept of neutrinospheres is perhaps an oversimplification of
the more
detailed statistics of neutrino emission as displayed in Fig. (2).
Neutrinospheres should not be energy averaged, and the neutrino emission
should not be based on a core diffusion scenario alone. Most of the
neutrinos are emitted from the base of the cooling region where infalling
material settles onto the surface of the protoneutron star. Moreover,
the separation into cooling and heating region refers to the symptom
instead of the cause. A separation into regions of thermal balance,
weak interaction equilibrium, and non-equilibrium in the
spirit of \cite{Herant_et_al_94} could be more appropriate. The
region where matter is in thermal balance and weak equilibrium with the
(non-equilibrium!)
background neutrino field, is congruent with the former cooling region.
Additional energy deposition cannot
be achieved by prolonged exposure, only by a change of the matter
density or the background
neutrino luminosities and rms energies. In Fig. (3a),
\begin{figure}[t]
\includegraphics[width=0.9\textwidth]{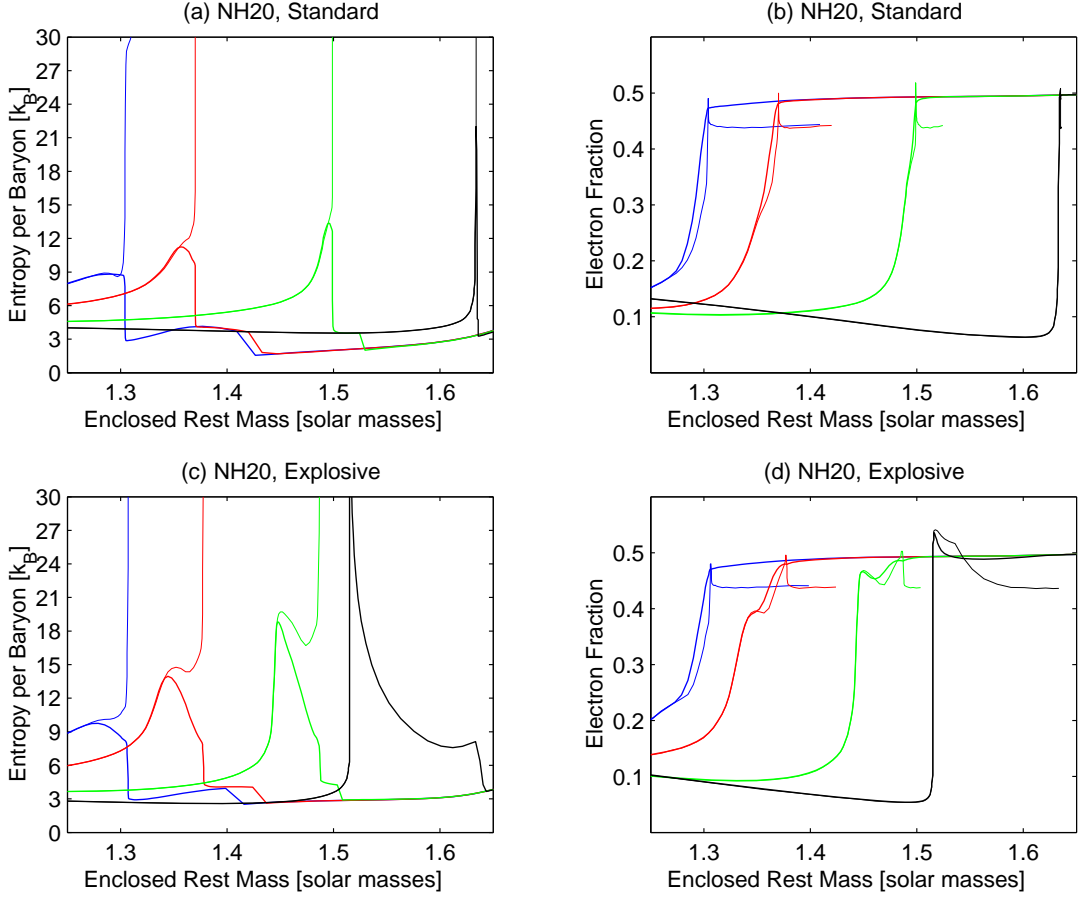}
\caption{Each graph in this figure shows, from left to right, four snapshots
of entropy (a,c) or electron fraction (b,d) profiles as a function of mass at
\protect\( 50 \protect\), \protect\( 100 \protect\),
\protect\( 200 \protect\), and \protect\( 450 \protect\) ms after bounce.
The profiles
found in the evolution of the full run are represented with a thick line.
The thin line shows the
asymptotic values, obtained by infinite exposure time to the prevailing neutrino
abundances.}
\end{figure}
we compare entropy profiles in the evolution of the \( 20 \) M\( _{\odot } \)
progenitor to equilibrium values. We note that
at \( 50 \) ms after bounce infalling matter reaches thermal balance
immediately after shock passage. There is no neutrino heating possible.
At \( 100 \) ms after bounce, the equilibrium entropy is higher than
the value reached by shock dissipation. Inflowing material joins
the equilibrium line at the former gain radius
and a negative entropy gradient becomes manifest. It is, however,
counter-intuitive to imagine it as the source of the extremely
vigorous convection
and fast shock expansion found in multi-dimensional simulations of
the last decade. After \( 200 \) ms, the failure of the model to
produce an explosion becomes evident. A similar analysis can be made for
the electron fractions shown in graph (b).
Infalling matter decreases the electron
fraction and increases the entropy. Graphs (a) and (b) also illustrate
the possible gain in heating efficiency with convection.
Models with stratified hydrodynamics suffer from the fact
that the equilibrium entropy is reached in the shells best
exposed to neutrino absorption, and that further out, where higher
entropies would be allowed, the rates are too low to approach equilibrium
during infall time \cite{Herant_et_al_94}.
Convection can mediate in this situation by
exchanging material between these two domains such that more material
is heated towards equilibrium entropy. It
is interesting that the equilibria provide
a quantifiable upper and lower bound respectively for entropy and
electron fraction changes in the heating region.

\section{Sensitivity to nuclear and weak interaction input physics}

The failed supernova models in spherical symmetry have, of course,
triggered many suggestions to improve the input physics in
hopes of overcoming the dead end. One can
distinguish improvements that affect the initial bounce-shock
energy from improvements that influence the neutrino-driven shock
revival. The prompt shock fails in simulations with detailed
neutrino transport \cite{Myra_Bludman_89} and loses \emph{all} of its
kinetic energy in the neutrino burst long before neutrino heating.
Unless there are \( \sim 40\% \) changes in the enclosed mass
at shock formation, the influence of the bounce-shock energetics to
the later shock revival by neutrino heating is marginal.
Most interesting for the neutrino-driven mechanism are thus improvements
in the standard input physics that alter neutrino
luminosities and rms energies. We name some possibilities from inside
out: Changes in the high density equation of state would change the
compactness of the protoneutron star and with it the position of the
accretion layers in the gravitational potential. A more compact protoneutron
star would lead to larger infall velocities, higher neutrino luminosities,
and higher rms energies in analogy to the impact of general
relativistic effects in the core \cite{Bruenn_DeNisco_Mezzacappa_01}.
Moreover, uncertainties are attributed to the neutrino opacities
at high densities \cite{Burrows_Sawyer_98} and corrections
for correlations and nucleon recoil should be implemented. However,
we speculate that the early luminosities would not dramatically change
because the affected density range is enclosed in layers of subnuclear
densities, whose still appreciable opacities would prevent dramatic
boosts in the neutrino outflow during the first few hundred milliseconds
after bounce. Note that the maximum density at
\( 100 \) ms after bounce is only about twice the saturation density.
Additional sources for the \( \mu  \) and \( \tau  \) neutrinos
have been suggested: production by bremsstrahlung \cite{Thompson_Burrows_Horvath_00} 
and production by electron neutrino pair annihilation \cite{Keil_Raffelt_Janka_02}.
Both reactions have been found to locally significantly exceed the
importance of the traditional electron-positron pair process. However,
changes in self-consistent simulations are small, i.e.
on the \( 10\% \) scale in \( \mu/\tau \) neutrino luminosities
\cite{Keil_Raffelt_Janka_02}. Around the neutrinospheres,
a luminosity boost could occur if an equation of state would lead to
convective instabilities.
However, a stability analysis of the Lattimer-Swesty equation of state
in spherical symmetry does not indicate instabilities \cite{Wilson_Mayle_93}.
Opacity improvements at lower densities are relevant in the early
postbounce evolution, because they directly influence the radiation
hydrodynamics at and above the surface of the protoneutron star. A
list of possible corrections has been presented in \cite{Horowitz_02}.
Corrections for weak magnetism and nucleon recoil have been investigated
\cite{Rampp_Janka_02} and corrections for the strangeness content
of nucleons have been explored \cite{Liebendoerfer_et_al_02}. In
Fig. (4a),
\begin{figure}[htb]
\includegraphics[width=\textwidth]{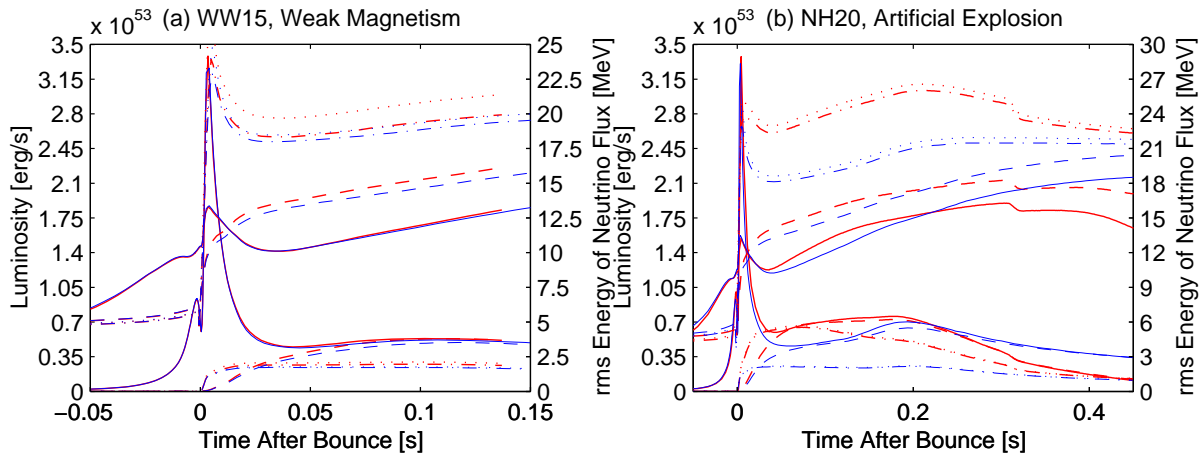}
\caption{Graph (a) investigates the effect of weak magnetism on electron
neutrinos (solid), electron antineutrinos (dashed), \protect\( \mu/\tau
\protect\) neutrinos (dash-dotted), and \protect\( \mu/\tau \protect\)
antineutrinos (dotted). The improved luminosities and rms energies (thick
lines) are compared to standard weak interactions (thin) lines. The same
neutrino coding is used in graph (b), where we
compare the luminosities and rms energies
in an artificial explosion (thick lines) to the standard case (thin lines).
}
\end{figure}
we show the effect of strangeness and weak magnetism
on the luminosities and rms energies in a general relativistic evolution
of the \( 15 \) M\( _{\odot } \) progenitor. Note that, in all three
flavors, neutrinos are affected differently than antineutrinos. However,
the influence of these corrections on the hydrodynamics are barely
visible in our simulation.

\section{Phenomenology of neutrino driven explosions in spherical
symmetry\label{section_explosion_phenomenology}}

The suggested improvements in input physics are quite far from leading
to neutrino-driven explosions in the spherically symmetric models.
Nevertheless, it is interesting to change input physics beyond physical
limits to explore the phenomenology of neutrino-driven explosions with
detailed neutrino transport. Explosions can be obtained by the combination
of the following measures: (a) The scattering of
neutrinos on free nucleons is parameterized to 40\% of the full
cross section. (b) Only two angles (60, 120 degrees from radius) are
considered for the neutrino transport, this increases the heating
efficiency and the computing efficiency
for parameter studies. (c) The diffusion limit is finite differenced
with interpolated transport coefficients (see \cite{Mezzacappa_Messer_99}
appendix C for a discussion).
The exploding models do still obey lepton number and energy conservation
and can be claimed self-consistent in this respect. The luminosities
of the artificially exploding model are compared to the luminosities
of the standard model in Fig (4b). The heavy neutrinos most
effectively reflect the enforced reduction in neutral current reactions.
Reduced scattering on the way out reduces the effective path length
subject to neutrino electron scattering. The neutrinos cease to thermalize
earlier and escape with higher rms energies. Measurements
of heavy neutrino properties from supernovae may best single out information
about scattering opacities in the protoneutron star. The electron
flavor luminosities are about \( 30\% \) higher than standard during
the first \( 150 \) ms and drop to \( 50\% \) lower values later
because of the reduced accretion luminosity after the launch of the
explosion. Equipartition of the luminosity across the three flavors
sets in when the accretion becomes negligible. The equilibrium constraints
in Fig. (3c) show that already at \( 50 \) ms after bounce
the upper entropy bound is much higher than in the standard case,
graph (a). As the evolution
continues, the neutrino properties are always such that there is ample
room for entropy increase. As the explosion proceeds, a large negative
entropy gradient develops across a mass of order \( 0.1 \) M\( _{\odot } \),
in good agreement with analytical estimates \cite{Janka_01}. These are future
ejecta from the silicon and oxygen layers. All our artificially exploding
models produce an electron fraction above \( 0.5 \) in the neighborhood
of the masscut. As the material
at the mass cut is heated and expanded, the electrons become non-degenerate
and the electron chemical potential \( \mu _{e} \) falls below the
neutron to proton mass difference \( Q \). Rising electron fractions
approaching \( 0.5 \) have already been predicted in the protoneutron
star wind \cite{Qian_Woosley_96}. Here in our high entropy bubble, the
tighter bound protons become even more abundant than neutrons as soon
as \( \mu _{e}<Q/2 \) \cite{Beloborodov_02}. Graph (d) demonstrates that
material around the masscut actually reaches this weak equilibrium
at late times.

\footnotesize

\section*{Acknowledgements}

We acknowledge funding by the NSF under contract AST-9877130, the
Oak Ridge National Laboratory, managed by UT-Batelle, LLC, for the
U.S. Department of Energy under contract DE-AC05-00OR22725, the Swiss
National Science Foundation under contract 20-61822.00, and the
DoE HENP SciDAC Program. Our Simulations have been carried out on
the NERSC Cray SV-1.

\end{document}